\documentclass[11pt]{article}
\begin{document}

\newcommand{\be}{\begin{equation}}
\newcommand{\ee}{\end{equation}}
\newcommand{\rf}[1]{(\ref{eq:#1})}

\title{ON THE NATURE OF THE COSMOLOGICAL  CONSTANT PROBLEM}

\author{M. D. MAIA\footnote{maia@unb.br}
\& A. J. S.CAPISTRANO\footnote{capistranoaj@unb.br}\\
Instituto  de F\'\i sica, Universidade de Bras\'\i lia,
  Bras\'\i lia D.F.  70919-970\\
E. M.  MONTE\footnote{edmundo@fisica.ufpb.br}\\
Departamento de F\'\i sica,  Universidade Federal da Para\'\i ba, Jo\~ao Pessoa, 58059-970}

\maketitle

 \begin{abstract}

General  relativity postulates the Minkowski  space-time  to be the   standard flat geometry  against  which  we compare  all  curved  space-times and  the gravitational ground  state where  particles, quantum  fields  and  their vacuum   states are primarily conceived. On the other hand,   experimental  evidences   show  that  there exists  a  non-zero cosmological  constant,  which implies in a  deSitter space-time,   not   compatible  with the assumed Minkowski structure.  Such  inconsistency   is  shown  to be   a  consequence of  the lack of a application independent curvature  standard  in  Riemann's  geometry,   leading eventually to  the  cosmological constant problem in general  relativity.

We  show how  the  curvature standard in Riemann's  geometry  can  be  fixed  by    Nash's  theorem on locally  embedded  Riemannian  geometries, which imply  in the existence of  extra  dimensions.  The resulting gravitational  theory  is  more general than  general  relativity,   similar  to   brane-world gravity, but  where the  propagation of the gravitational field along the extra  dimensions is  a mathematical necessity,   rather  than being a a postulate.   After  a  brief introduction to  Nash's  theorem,  we show  that   the    vacuum energy density must  remain confined  to four-dimensional  space-times, but  the   cosmological  constant resulting from the  contracted  Bianchi  identity is  a gravitational contribution  which propagates  in the extra  dimensions.   Therefore,   the comparison  between the vacuum  energy and   the  cosmological  constant in  general relativity ceases  to be. Instead,    the geometrical  fix  provided by   Nash's  theorem  suggests that  the   vacuum  energy density  contributes    to   the  perturbations of  the  gravitational  field.

\end{abstract}

\section{The Cosmological Constant Problem}

Back in the  60's,  Zel'dowich   showed  how  the fluctuations of  quantum  fields  can be described  as   a  perfect fluid  with  state   equation  $p_v = -<\rho_v>=$constant. Then,  using  the semi-classical  Einstein's  equations,  it was shown  that such  fluid adds  a   non-trivial contribution  to   the gravitational  field  as
\be
 R_{\mu\nu} -\frac{1}{2}R g_{\mu\nu} +\Lambda g_{\mu\nu}=  8\pi  G T_{\mu\nu}^m  +
8\pi G <\rho_v>g_{\mu\nu} \label{eq:EEvac}
\ee
where  $T_{\mu\nu}^m$  denotes  the   energy-momentum tensor of the classical  sources  \cite{Zeldowich}.
Comparing the  constant  terms in both sides of this  equation we  obtain  $ \Lambda/8\pi G     = <\rho_v>$, or  as it is commonly   stated, \emph{``the  cosmological constant is the  vacuum  energy density"}.
However,   such  conclusion still depends on the solution of the cosmological constant problem:  Current observations tell that  $\Lambda/8\pi G  \approx  10^{-47} \;Gev^4$  ($c=1$),  while  the  estimates  of the vacuum energy density is   $<\rho_v>  \approx 10^{76}\; Gev^4$.
It is  somewhat  disappointing  that this  difference
cannot be resolved by  any  known theoretical   procedure in  quantum  field  theory \cite{Weinberg}.

This problem is  related to the  fact that  Riemann's  geometry  describes  classes   of  curvature  equivalent   manifolds, defined by  the   same  curvature  tensor $R(U,V) W  = \nabla_U \nabla_V W -\nabla_V \nabla_U W  -\nabla_{[U,V]}W $ defining    the  local shape of  a manifold, without  reference   to any  previously  established   standard  of  shape or  curvature. Riemann's   own   example  is  given    by  a flat    manifold  defined  by   $R(U,V)W=0$  which  can  be   a  plane,  a cylinder,  in fact  any  ruled  manifold\cite{Riemann}. A  solution of such  ambiguity was conjectured without proof by L.  Schlaefli  in 1873,  suggesting   that   any Riemannian  manifold  should  be   embedded into  a larger flat  manifold,  acting as  the curvature   reference\cite{Schlaefli}. The  final  proof   of
 that   conjecture  based on metric perturbations was  accomplished by   J.  Nash in     1956.

The ground  state  of the   gravitational  field in general  relativity  is  the   Minkowski  tangent  space-time, taken  as   the flat-plane standard   of  curvature,  in  which   particles, quantum  fields,   their vacuum sates  and energy  are conceived.  However,   Minkowski's  geometry owes its plane flatness  to  the  Poincar\'e  symmetry of  Maxwell's  equations, and  not to Riemann's geometry.  On the other hand,    modern   cosmological  observations indicate  that the  cosmological  constant  $\Lambda$  albeit small is  not  zero. The existence of such constant requires  that  the  quantum  fluctuations of the vacuum  should be defined in a  deSitter ground  state instead  of  the  Minkowski one.
Therefore, the cosmological  constant problem  results  from the fact that the  class  of  equivalence of  manifolds which  contain  $\Lambda$ is  not the  same  class of  equivalence  of  manifolds  which  contain  the flat   Minkowski's  space-time.

In the  following  section we apply  Nash's solution of    Schlaefli's  conjecture   to   gravitational   physics,
showing  why    the   emergence of   extra  dimensions  is necessary to remove  Riemann's  curvature ambiguity.  In section 3   we  apply this  result   to the   cosmological  constant problem.

\section{Embedded Riemannian Structures}

Consider  a  D-dimensional Riemannian    manifold   $V_D$   with   metric  geometry  ${\cal G}_{AB}$,  and     another    Riemannian  manifold  $V_n$,  $n<D$    with  metric $g_{\mu\nu}$. The local and isometric  embedding  of $V_n$    in  $V_D$  is  the  map
$\mathcal{Z}:  V_n \rightarrow V_D$    with $D$ components
 $\mathcal{Z}^A (x^\mu,y^a)$   functions of the coordinates  $x^mu$ in $V_n$  and the extra  coordinates $y^a$  orthogonal to $V_n$, such  that  (Index  convention:  $\mu, \nu
=1\cdots n,  \;\;  a,b =n+1\cdots D,\;\;  A,B =1\cdots D $)
\[
\mathcal{Z}^A_{,\mu}\mathcal{Z}^B_{,\nu}\mathcal{G}_{AB}
=g_{\mu\nu},\;\;\mathcal{Z}^A_{,\mu}\eta^B_{a}\mathcal{G}_{AB}=0,\;\;
\eta^A_{a}\eta^B_{b}\mathcal{G}_{AB}=g_{ab}
\]
where  $\left\{\eta_a\right\}$ is   a  basis  of  the  $N=D-n$-dimensional complementary   space,   orthogonal    to  $V_n$.  With this  we may  construct a  Gaussian frame of  $V_D$ based on $V_n$ $\{\mathcal{Z}^A_{,\mu}, \eta^A_a\}$.
The   embedded  Riemannian  geometry differs  from the  non-embedded  one  in  that  besides the metric
there are  two  additional variables.  The  extrinsic  curvature
defined by    $k_{\mu\nu a}=  -\mathcal{Z}_{,\mu}^A  \eta^B_{,\nu}\mathcal{G}_{AB}$ and  the   third  fundamental  form  defined by   $A_{\mu ab}=\eta^A_a  \eta^B_{b,\nu}\mathcal{G}_{AB}$.

Nash's  theorem  starts   with  the  analysis of smoothing  operators  on manifolds\cite{Nash},  specifying  that
 the embedding  map  must be  differentiable and  regular.
The  second and most important  part  of the  theorem tells   how     to generate  a  Riemannian  manifold   by    an   infinitesimal  deformations (or perturbations)  of $V_n$  along the extra  dimensions given by
\be
\delta \bar{g}_{\mu\nu}=-2\bar{k}_{\mu\nu  a }\delta y^a
\label{eq:pertu}
\ee
Therefore,  for  a  given metric   $\bar{g}_{\mu\nu}$, we obtain a perturbed  metric  $g_{\mu\nu}= \bar{g}_{\mu\nu}  +\delta  \bar{g}_{\mu\nu}$
and  by  a  successive  repetition of this procedure,
any  differentiable embedded  Riemannian  manifold may  be  generated. Assuming that  the embedding is  regular, then   we may
use  the inverse function  theorem  to   un-embed  the  perturbed   geometry,    obtaining a   purely  intrinsic  Riemannian  geometry.
(More  details on Nash's  theorem and its  applications to cosmology and  quantum gravity can be found in\cite{Maia}.).

The  4-dimensionality of  the  space-time is   a  direct  consequence of  the   Hodge duality of the  gauge  fields:  Denoting
$F=F_{\mu\nu}dx^\mu \wedge dx^\nu,\;\;
F_{\mu\nu}=  [D_\mu , D_\nu ], \;\; D_\mu =\partial_\mu + A_\mu$,
then  the Yang-Mills   equations can be written as
$  D \wedge F=0\;\;  \mbox{and}\;\;    D\wedge F^*  =4\pi j^*$
where $D =D_\mu dx^\mu$  and $F^*$  is  the  dual of  $F$:
$ F^*   =F^{*}_{\mu\nu} dx^\mu \wedge dx^\nu,\;\; F^*_{\mu\nu} =\epsilon_{\mu\nu\rho\sigma}F^{\rho\sigma}$.
Such  duality  holds only   when  a  3-form is isomorphic  to a 1-form,  which is typical of  four  dimensional  space-times.
On the other hand,  \rf{pertu} implies that  the  gravitational  field  propagates  along  the  extra  dimensions of  the  embedding  space.

In principle, the  embedding  space    can  be  any  Riemannian  manifold.  However,  the differentiable   condition required by  Nash's  theorem  suggests  that the geometry of that space must obey the  Einstein-Hilbert  principle,   stating that  the  metric  geometry must be  the  smoothest possible.
Consequently,  the reference  geometry (the bulk geometry) is   defined by the local Einstein's  equations
 \be
{\mathcal{R}}_{AB}-\frac{1}{2}\mathcal{R} \mathcal{G}_{AB}
=\alpha_* T^*_{AB}, \;\;\;   A,B =  1 \cdots D     \label{eq:EEbulk}
\ee
 where  $\alpha_*$ denotes  the $D$-dimensional  energy  scale
 and   $T^*_{AB}$  is    the  energy-momentum  tensor  of  the known  sources,  which  we assume  to be composed of  4-dimensional  confined   fields  and ordinary  matter. In this case,  the  confinement  can  be generically expressed   as
$
\alpha_* T^*_{\mu\nu}=8\pi G T_{\mu\nu},
 \;\;\; T^*_{\mu a}=T^*_{ab} =0.
$

In  the  following   we  will consider the simpler case of  a  flat   5-dimensional embedding  space   such that  $ \mathcal{R}_{ABCD}=  0$. This is    sufficient  to embed  the  standard  cosmological model\cite{Maia} and it is  sufficient  to our  analysis of the cosmological constant problem.
 The  gravitational  equations  for the embedded  geometry  are  obtained   the equations \rf{EEbulk} written  in the mentioned Gaussian   frame
 \begin{eqnarray}
&&R_{\mu\nu}-\frac{1}{2}Rg_{\mu\nu}+\Lambda g_{\mu\nu} +
{Q}_{\mu\nu} = 8\pi G T_{\mu\nu},\;\;\;
k_{\mu ;\rho}^{\rho}\!
 -\!h_{,\mu} = 0  \label{eq:BE1}
\end{eqnarray}
 where we  have denoted   $k_{\mu\nu 5}=k_{\mu\nu}$,   $h= g^{\mu\nu}k_{\nu\nu}$,    $K^{2}=k^{\mu\nu}k_{\mu\nu}$ and  the
conserved  extrinsic  geometric term
\begin{equation}
Q_{\mu\nu} = -k^{\rho}{}_{\mu }k_{\rho\nu }+ h
k_{\mu\nu}\!\!  +  \!\!\frac{1}{2}(K^{2}-h^{2})g_{\mu\nu},
\;\;\;Q^{\mu\nu}{}_{;\nu} =0   \label{eq:Qij}
\end{equation}
which is   a  consequence of the embedding  equations  \cite{Maia}.  The  cosmological   constant $\Lambda$ appears  after  the  contracted  Bianchi identity.

Now,   we  may  return  to the    cosmological constant   problem.
Assuming again the  validity  of the semi-classical  regime   for the contribution of     the  quantum fluctuations of  the confined vacuum,  then  equation \rf{BE1}  becomes
 \begin{eqnarray}
&&R_{\mu\nu}-\frac{1}{2}Rg_{\mu\nu}+\Lambda g_{\mu\nu} +
{Q}_{\mu\nu} = 8\pi G T^m_{\mu\nu} + 8\pi G
<\rho_v>g_{\mu\nu} \label{eq:BEV1}
%\\
% k_{\mu ;\rho}^{\rho}\!
% -\!h_{,\mu} = 0\label{eq:BEV2}
\end{eqnarray}
However, in contrast with the case of  general  relativity,
 $\Lambda$ and  $<\rho_v>$ have  different
 meanings and  dynamical  behavior:  While $\Lambda$ represents  a direct contribution  to the gravitational  field,  the vacuum energy   $<\rho_v>$ remains   confined.  Furthermore,  except in the cases  where   $k_{\mu\nu}=0$,  it is  not possible  to  cancel $\Lambda/8\pi G$ with $<\rho_v>$, because $Q_{\mu\nu}$  is  a function of $k_{\mu\nu}$, so that it also  contributes  to the propagation of the gravitational field   in the extra  dimensions according to  \rf{pertu}.

The overall  conclusion is  that  the  cosmological  constant problem is proper of general  relativity,  which inherits the ambiguity associated  with  the equivalence classes  of Riemann's curvature. Either   we  take   the    Minkowski's ground  state
as the standard flat  plane,  where    $<\rho_v>$  is  defined,  or  else, in face of the observations of  a  non-zero   $\Lambda$,
 we take  the deSitter  ground state.   The  solution of  such  conflict comes  from the   fix  to Riemannian geometry   suggested by Schlaefli and demonstrated by Nash,  placing the   curvature   standard  in the bulk geometry defined by the Einstein-Hilbert principle.  The presence of  this higher-dimensional embedding  space provides  a  geometric  standard  of  curvature, making it possible  to contemplate  the  confined   vacuum structure and   the  deSitter space-time   without  the  cosmological  constant conflict.

\end{document}